\documentclass[aps,twocolumn,nofootinbib,groupedaddress,superscriptaddress,longbibliography]{revtex4-2}

\usepackage{amsmath,amssymb}
\usepackage{graphicx}
\usepackage{multirow}
\usepackage{bm}
\usepackage{mathtools}
\usepackage{dsfont}
\usepackage{amsfonts}
\usepackage{xcolor}
\usepackage[normalem]{ulem}
\usepackage{url}

\usepackage[colorlinks=true,linkcolor=magenta,citecolor=magenta,urlcolor=magenta,hypertexnames=false]{hyperref}

\def\ba#1\ea{\begin{align}#1\end{align}}
\def\bg#1\eg{\begin{gather}#1\end{gather}}
\def\bpm{\begin{pmatrix}}
\def\epm{\end{pmatrix}}
\newcommand{\nn}{\nonumber \\ }
\newcommand{\bb}[1]{{\mathbf #1}}
\newcommand{\bx}{\bb x}
\newcommand{\bk}{\bb k}
\newcommand{\bR}{\bb R}
\newcommand{\mc}[1]{\mathcal{#1}}
\newcommand{\mf}[1]{\mathfrak{#1}}

\newcommand{\td}[1]{\widetilde{#1}}

\newcommand{\der}{\partial}
\newcommand{\sgn}{{\rm sgn}}
\newcommand{\dg}{\dagger}

\newcommand{\sg}{\sigma}
\newcommand{\vph}{\varphi}

\newcommand{\ep}{\epsilon}
\def\PRLgreater{\,{>}\,}
\def\PRLless{\,{<}\,}
\def\PRLequal{\,{=}\,}
\def\PRLminus{\,{-}\,}
\def\PRLplus{\,{+}\,}

\allowdisplaybreaks

\begin{document}
\title{Spin-momentum Locking and Topological Vector Charge Response with Conserved Spin}

\author{Yoonseok Hwang}
\thanks{These two authors contributed equally.}
\affiliation{Department of Physics, University of Illinois Urbana-Champaign, Illinois 61801, USA}
\affiliation{Anthony J. Leggett Institute for Condensed Matter Theory, University of Illinois Urbana-Champaign, Illinois 61801, USA}

\author{Penghao Zhu}
\thanks{These two authors contributed equally.}
\affiliation{Department of Physics, University of Illinois Urbana-Champaign, Illinois 61801, USA}
\affiliation{Anthony J. Leggett Institute for Condensed Matter Theory, University of Illinois Urbana-Champaign, Illinois 61801, USA}

\author{Taylor L. Hughes}
\affiliation{Department of Physics, University of Illinois Urbana-Champaign, Illinois 61801, USA}
\affiliation{Anthony J. Leggett Institute for Condensed Matter Theory, University of Illinois Urbana-Champaign, Illinois 61801, USA}

\begin{abstract}
Spin-momentum locking plays a fundamental role in spintronics and, more broadly, is an important concept in condensed matter physics.
In 2D and 3D, spin-momentum locking typically does not allow spin conservation because the  spin-1/2 operators of electrons anticommute.
Instead, here we study spin-momentum locking terms with conserved, commuting pseudospins built from a combination of spin and orbitals.
We find that 2D spin-momentum locking terms with conserved pseudospins generally lead to linearly dispersing modes at low-energy with anomalous charge and pseudospin currents.
To cure the anomaly we show that such anomalous modes can be realized on the surface of a 3D Weyl semimetal (or an associated weak topological insulator) with a nonzero mixed spin-momentum quadrupole moment, which is determined by the momentum location and pseudospin eigenvalues of Weyl points at the Fermi level.
Crucially, this mixed quadrupole moment captures a mixed pseudospin-charge bulk response that cancels the anomaly of surface modes, and can generate a giant 3D spin Hall effect, among other phenomena.
\end{abstract}

\maketitle

\section{Introduction}
Spintronics, and in general the electrical control of spin, is built upon the relativistic phenomenon of spin-orbit coupling (SOC)~\cite{dresselhaus1955spin,bychkov1984properties,bychkov1984oscillatory, winkler2003spin,manchon2015new,soumyanarayanan2016emergent}.
In solids, one remarkable, low-energy manifestation of atomic SOC is the concept of spin-momentum locking (SML), where the orientation of an electron's spin is inherently tied to its direction of motion.
SML is the foundation for many of the most exciting spintronic phenomena~\cite{vzutic2004spintronics} in two-dimensional (2D) electron systems with Rashba SOC~\cite{bychkov1984oscillatory,bychkov1984properties}, the 2D surface states of topological insulators (TIs) ~\cite{liu2010model,qi2010topological}, and the 1D helical edge states of quantum spin Hall insulators~\cite{kane2005quantum,bernevig2006quantum,wu2006helical,xu2006stability,roushan2009topological}.

Previous studies of SML have focused on couplings between momenta and the intrinsic spin-1/2 of the electrons.
Because of the anticommuting nature of the spin-1/2 Pauli matrices, spin is not conserved in 2D and 3D systems exhibiting SML as more than one Pauli matrix generally appears in the low-energy Bloch Hamiltonian.
This is not ideal for spintronics applications as the nonconserved spins have a finite lifetime and any stored information will decay as the spins undergo unconstrained equilibration.
With this problem in mind, it is therefore crucial to ask if SML and conserved spins can coexist.

Below we demonstrate that the answer to this question is yes, and that these systems can be anomalous in the sense that external fields induce a mixed spin-charge 't Hooft anomaly in some cases.
This anomaly dictates both the conditions under which such systems exist in their intrinsic spatial dimension, and their response properties to external fields when the anomaly is remedied by attaching the system to the boundary of a higher-dimensional bulk.
Importantly, we find that in some crystal symmetry classes, the higher dimensional bulk response that compensates for the surface anomaly is captured by an inherently non-relativistic, 3D rank-2 symmetric or antisymmetric axial vector charge Chern-Simons term, where the axial-vector charge is the conserved spin.
Interestingly, we show that the bulk response is naturally generated by a Weyl semimetal with a nodal configuration that generates a mixed spin-momentum Weyl quadrupole.
Here, the quadrupole moment refers to a weighted sum of Weyl node positions multiplied by their pseudospin eigenvalues, analogous to how momentum-space dipole moments determine anomalous Hall responses in ordinary Weyl semimetals.
This is in analogy to the recently proposed rank-2 chiral fermions which carry a crystal momentum vector charge and a momentum-momentum Weyl quadrupole~\cite{dubinkin2024higher,zhu2023higher}.

Explicitly, here we study 2D SML terms in {\it nonmagnetic} crystalline systems with rotation and mirror symmetries.
We focus on effective $k\cdot p$ Hamiltonians having the form
\bg
h_\bk = \sum_{i = x,y} \sum_{A = X,Y} g_{iA} k_i S_A,
\label{eq:surface_general}
\eg
where $(S_X,S_Y)$ are spin operators satisfying $S_A^\dg \PRLequal S_A$, and ${\rm Tr}[S_A] \PRLequal 0$, and $g_{iA}$ is $2 \times 2$ matrix with real entries that will be constrained by the imposed symmetries.
Note that we work in units $\hbar \PRLequal c \PRLequal 1$ and the electron charge is $\PRLminus e$.
We expect that effective Hamiltonians of this form may appear when the full Bloch Hamiltonian is expanded around time-reversal invariant momenta (TRIM) in the Brillouin zone (BZ).
In the neighborhood of a TRIM, time-reversal symmetry (TRS) simply flips both spin and momentum in Eq.~\eqref{eq:surface_general}, and hence the effective Hamiltonian is manifestly time-reversal invariant.

The simplest case of SML with conserved spin in Eq.~\eqref{eq:surface_general} is when only one spin operator appears.
This happens in the 1D helical edge states of the quantum spin Hall insulator, but in 2D, this case is allowed only in lower-symmetry systems having 2D point groups $C_{1,2}$ or $C_{1v,2v}$ (and even then would require extreme anisotropy or fine-tuning)~\cite{winkler2003spin,samokhin2022effective}.
We note that for $C_3$ and $C_{3v}$ symmetries, there can also be SML terms with one conserved spin operator $S_Z$, which are cubic in momentum~\cite{fu2009hexagonal,samokhin2022effective}, but we leave further analysis of higher-order terms as future work.

Our primary focus will instead be on systems with $C_{3,4,6}$ or $C_{3v,4v,6v}$, where the SML terms that are {\it linear} in momentum always involve both $S_X$ and $S_Y$, as shown in Sec.~S1 of the Supplementary Material (SM)~\cite{supple}.
In these cases the spin is conserved only when $S_X$ and $S_Y$ commute with each other, i.e., $[S_X,S_Y] \PRLequal 0$.
Although the intrinsic spin-1/2 of the electrons cannot satisfy this condition on their own, we will show that commuting operators consisting of both orbital and spin degrees of freedom of electrons can be constructed.
Hence, we study generalized SML terms in Eq.~\eqref{eq:surface_general} with commuting pseudospin operators $S_X$ and $S_Y$ that are both conserved.

\section{Rashba term with commuting pseudospins}
For our primary example we focus on the point group $C_{4v}$, which is promoted to the space group $P4mm$ for lattice Hamiltonians.
We further limit our study to nonmagnetic systems having TRS $\mc{T}$ with $\mc{T}^2 \PRLequal \PRLminus 1$.
The relevant symmetry elements in $C_{4v}$ act on the position-space coordinates as $M_x: (x,y,z) \to (\PRLminus x,y,z)$, $M_y: (x,y,z) \to (x, \PRLminus y,z)$, and $C_{4z}:(x,y,z) \to (\PRLminus y,x,z)$.
Their representation is chosen as
\bg
M_{x,y} = - i \tau_0 \sg_{x,y}, \,
C_{4z} = \tau_z e^{- i \frac{\pi}{4} \sg_z}, \,
\mc{T} = - i \tau_0 \sg_y \mc{K},
\label{eq:symmop}
\eg
where $\mc{K}$ is complex conjugation, $\sg_a$ are the Pauli matrices for the spin-1/2 degree of freedom, and the $\tau_a$ Pauli matrices represent a pair of orbitals transforming under the trivial (e.g., $s$ orbital) and $B_1$ (e.g., $d_{x^2 \PRLminus y^2}$ orbital) representations of $C_{4v}$.
From these degrees of freedom we can construct commuting pseudospin operators $S_{X}$ and $S_{Y}$:
\bg
S_X = \frac{1}{\sqrt{2}} (\tau_z + \tau_x) \sg_x,
\quad S_Y = \frac{1}{\sqrt{2}} (\tau_z - \tau_x) \sg_y.
\label{eq:spins}
\eg
Then, $S_X$ and $S_Y$ commute with each other and transform under $C_{4v}$ and $\mc{T}$ as a axial vector: $M_x$ ($M_y$) flips $S_Y$ ($S_X$), $\mc{T}$ flips both $S_X$ and $S_Y$, and $C_{4z}$ rotates $(S_X,S_Y)$ to $(S_Y,-S_X)$.

\begin{figure}
\centering
\includegraphics[width=0.48\textwidth]{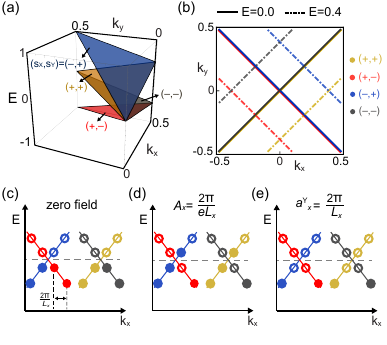}
\caption{
Energy spectrum and anomalous responses of $h_{\bk}$.
(a) A portion of the energy spectrum for eigenstates of $h_{\bk}$ with different pseudospins $(s_X,s_Y)$.
(b) FS contours of $h_{\bk}$.
[(c) and (d)] Low-energy spectra and fillings as a function of $k_x$ at a fixed $k_y \PRLgreater 0$ when (c) the gauge fields are zero, (d) the charge $U(1)_{\rm EM}$ gauge field is adiabatically shifted by $2\pi/(eL_x)$, and (e) the spin $U(1)_Y$ gauge field is adiabatically shifted by $2\pi/L_x$.
Filled (empty) circles represent occupied (unoccupied) states.
The pseudospin sectors $(\PRLminus,\PRLplus)$, $(\PRLplus,\PRLplus)$, $(\PRLplus,\PRLminus)$, and $(\PRLminus,\PRLminus)$ are presented by blue, yellow, red, and gray, respectively.
}
\label{fig1}
\end{figure}

Using this construction we can now consider SML phenomena with commuting spins.
Independent of the spin commutation relations, $C_{4v}$ and $\mc{T}$ symmetries restrict the coefficients $g_{iA}$ in Eq.~\eqref{eq:surface_general} to $g_{xY} \PRLequal \PRLminus g_{yX} \PRLequal v$ and $g_{xX} \PRLequal g_{yY}\PRLequal 0$ (see SM Sec.~S1~\cite{supple}).
Without loss of generality, we set $v \PRLequal 1$ in the following unless otherwise specified.
Thus Eq.~\eqref{eq:surface_general} is reduced to
\bg
h_{\bk} = k_x S_Y - k_y S_X,
\label{eq:rashba}
\eg
which is an analog of the Rashba term~\cite{bychkov1984oscillatory} but with commuting, instead of anticommuting, spin operators.
For simplicity, we call the Hamiltonian described by Eq.~\eqref{eq:rashba} a commuting-spin Rashba (CSR) term.
The spectrum of the CSR term has four bands $E_{\bk} \PRLequal \pm k_x \mp k_y$ and they, and the FS contours of $h_{\bk}$, are shown in Figs.~\ref{fig1}(a)-(b).
We see that $h_{\bk}$ has independent symmetries $U(1)_A$ generated by $S_{A \PRLequal X,Y}$, and each eigenstate of $h_{\bk}$ can be labeled by the quantum numbers $(s_X,s_Y)$.
Hence, each of the four open FSes has a fixed spin polarization that determines the Fermi-velocity $v_{\bk} \PRLequal (s_Y,\PRLminus s_X)$.
In comparison, instead of having four separate FSes that transform into each other under $C_4$, the conventional Rashba FS is a single closed circle where the spin-polarization and group velocity continuously wind around the FS.

\section{Anomalous charge and pseudospin currents}
We will now show that the CSR Fermi surfaces (FSes) have a mixed anomaly between the electric charge $U(1)_{\rm EM}$ and the spin $U(1)_A$ symmetries.
To demonstrate this, we view the CSR mode as a family of 1D chiral modes propagating in the $x$ direction ($y$ direction) with chirality $\chi_x(s_Y) \PRLequal \sgn(s_Y)$ [$\chi_y(s_X) \PRLequal \PRLminus \sgn(s_X)$] parameterized by each $k_y$ ($k_x$).
For example, at a fixed value of $k_y$ there are four linearly-dispersing modes propagating in the $x$ direction with chirality $\sgn(s_Y)$, each with an effective chemical potential proportional to the fixed $k_y$.
Hence, the charge and pseudospin anomalies of the CSR modes can be straightforwardly determined from the anomalies of 1D chiral modes.

Hence, to determine the anomalies of the CSR modes, let us first recall the  anomaly of a 1D chiral mode with dispersion $E_{k_x} \PRLequal \chi_x k_x$.
In the CSR model, each fixed-$k_y$ slice behaves as an effective 1D chiral mode, so viewing the system as a family of 1D modes provides a natural way to compute its anomaly.
Let us turn on an electric field by adiabatically shifting the
vector potential, $A_x$, by $2\pi/(eL_x)$ over a time period $T$.
Explicitly, we have $A_x(t) \PRLequal 2\pi t/(eL_xT)$ and electric field $E_x \PRLequal \PRLminus 2\pi/(eL_x T)$ for $t \in [0,T]$.
As $t$ goes from $0$ to $T$, the number of occupied right-moving modes ($\chi_x \PRLequal \PRLplus 1$) increases by one, and the number of occupied left-moving modes ($\chi_x \PRLequal \PRLminus 1$) decreases by one.
To interpret this as an anomalous conservation law we define the charge density $J^0 \PRLequal \PRLminus\frac{e}{V} \sum_{n\in occ.} 1$ where $V$ is the volume (area in 2D, length in 1D) of the system.
Thus, during this process $\Delta J^0 \PRLequal \PRLminus e \chi_x/V$, and the corresponding anomalous conservation law is 
\bg
\der_\mu J^{\mu} = \chi_x \frac{e^2}{2\pi} E_x.
\label{eq:ca}
\eg

Analogously, we first consider the anomalies of the CSR mode induced by the $U(1)_{\rm EM}$ gauge field $A_i$ ($i \PRLequal x,y$).
The gauge field $A_i$ couples with the CSR mode as:
\bg
h_\bk[A] = (k_x + e A_x) S_Y - (k_y + e A_y) S_X.
\label{eq:h_EM}
\eg
Considering the CSR mode as a family of 1D chiral modes, we see that the net chiralities along both the $x$ and $y$ directions are zero because chiral modes with opposite pseudospin eigenvalues have opposite chiralities as required by $\mc{T}$.
Therefore, the electric field will not induce an anomaly for the electric charge.
However, we can define the pseudospin density $\mc{J}_A^0
\PRLequal \frac{1}{V} \sum_{n \in {\rm occ}} s_{n,A}$ where $A\PRLequal X,Y$ and $s_{n,A}$ is the pseudospin eigenvalue of the $n$th eigenstate.
Interestingly, we find that the pseudospin density changes by a nonzero amount during an adiabatic shift of $A_i$, because the modes with opposite chiralities also have opposite pseudospin eigenvalues.
As illustrated in Figs.~\ref{fig1}(c) and \ref{fig1}(d), for a shift $\Delta A_x \PRLequal 2\pi/(eL_x)$ we find  $\Delta \mc{J}_Y^0
\PRLequal \frac{1}{V} \sum_{|k_y| \PRLless \Lambda_y} 4 \,
\PRLequal \frac{4 \Lambda_y}{\pi L_x}$, and similarly, for $\Delta A_y \PRLequal 2\pi/(eL_y)$ we find $\Delta \mc{J}_X^0 \PRLequal \frac{1}{L_x L_y} \sum_{|k_x| \PRLless \Lambda_x} (\PRLminus 4) \PRLequal \PRLminus\frac{4 \Lambda_x}{\pi L_y}$.
Here, $L_i$ and $\Lambda_i$ are the linear system sizes and momentum cutoffs along the $i \PRLequal x,y$ directions.
These responses can be expressed by the {\it anomalous} conservation laws:
\bg
\der_\mu \mc{J}^\mu_X
= \frac{2e \Lambda_x}{\pi^2 } E_y, \quad
\der_\mu \mc{J}^\mu_Y
= - \frac{2e \Lambda_y}{\pi^2} E_x,
\label{eq:anomaly}
\eg
with the electric field $E_i\PRLequal \PRLminus \der_t A_i \PRLplus \der_i A_0$ ($i \PRLequal x,y$).\footnote{We use the mostly-plus metric such that $A^0 \PRLequal \PRLminus A_0$ and $A^i \PRLequal A_i$.}
Furthermore, because of the $C_4$ symmetry it is natural to choose  $\Lambda_x \PRLequal \Lambda_{y} \PRLequal \Lambda.$

As befitting a mixed anomaly, we now discuss the inverse effect where a pseudospin electric field leads to an electric charge anomaly.
We denote the pseudospin $U(1)_A$ gauge fields as $a^A_i$ ($A \PRLequal X,Y$),\footnote{One can understand $a^A_i$ as the pseudospin connection.} and couple them to $h_\bk$ via
\ba
h_\bk[a] = (k_x - S_A a^A_x) S_Y - (k_y - S_A a^A_y) S_X.
\label{eq:h_spin}
\ea
Note that $a^A_i$ always appears as $S_A a^A_i$, which indicates that
the modes with opposite pseudospins feel opposite signs of the potential.
If we adiabatically shift $a_x^X$ by $2\pi/L_x$, then each $k_y$ contributes, $\Delta J^0 \PRLequal \PRLminus e$ $[\PRLplus e]$ in sectors with pseudospin eigenvalues $(\PRLplus 1, \PRLminus 1)$ and $(\PRLminus 1,\PRLplus 1)$ [$(\PRLplus 1, \PRLplus 1)$ and $(\PRLminus 1,\PRLminus 1)$].
Therefore the shift of $a_x^X$ does not lead to charge anomaly, and similarly for $a_y^Y$.
However, as illustrated in Figs.~\ref{fig1} (c) and (e), shifting $a^Y_x$ or $a^X_y$, generates nonzero charge density.
These responses are encoded in the anomalous charge conservation law:
\bg
\der_\mu J^\mu = \frac{2e \Lambda}{\pi^2}(\mc{E}^X_y - \mc{E}^Y_x),
\label{eq:charge_anomaly_c4v}
\eg
where the spin electric field is $\mc{E}^A_i \PRLequal \PRLminus \der_t a^A_i \PRLplus \der_i a^A_0$.

While Eq.~\eqref{eq:charge_anomaly_c4v} describes a charge anomaly in response to an antisymmetric combination of pseudospin electric fields, we can also find cases generated by a symmetric combination.
For example, we can choose $C_{4z}$ rotation such that $C_{4z}: (x,y,z) \to (\PRLminus y,x,z)$ and $(S_X,S_Y) \to (\PRLminus S_Y,S_X)$, such that the pseudospin rotates opposite to our previous convention.
This leads to a CSR term $h_{\bk} \PRLequal v(k_x S_Y \PRLplus k_y S_X)$, which has its charge anomaly induced by the symmetric combination of pseudospin electric fields, i.e., $\mc{E}^X_y \PRLplus \mc{E}^Y_x$ (see SM Sec.~S4~\cite{supple}).
Similar symmetric vector charge anomalies have been discussed in the recently proposed rank-2 chiral modes with dispersion $E \PRLequal k_x k_y$, where the charge anomaly is induced by a symmetric strain field, and the vector charge is momentum instead of pseudospin~\cite{dubinkin2024higher,zhu2023higher}.

\section{Realization of the CSR on the surface of a 3D semimetal}
The anomalies discussed above imply that the CSR modes cannot exist in isolation in a 2D lattice system.
There are two natural ways to treat this issue: (i) introduce symmetry preserving terms that generate closed, nonanomalous FSes intrinsically in 2D, or (ii) realize the CSR on the surface of a 3D system such that the bulk cancels the anomalous current via the Callan-Harvey anomaly inflow~\cite{callan1985anomalies}.
For a more interesting resolution to the anomaly let us consider mechanism (ii).
[See SM Sec.~S3~\cite{supple} for the case (i)].

To realize the anomalous CSR mode as a boundary mode we consider a 3D tight-binding (TB) model,
\ba
H_{3D}(\bk)
&= \sin k_x \, \mu_x S_Y - \sin k_y \, \mu_x S_X  + \sin k_z \, \mu_y 
\nn
&+ (m - 1 + \cos k_x + \cos k_y + \cos k_z) \, \mu_z,
\label{eq:h3D}
\ea
which is an eight-band model expressed using three sets of Pauli matrices $\mu$, $\tau$, and $\sg$; the latter two have the same interpretation as in Eq.~\eqref{eq:spins}.
This model has space group $P4mm$ whose symmetry generators are $C_{4v} \PRLplus \mc{T}$ and the 3D translations.
In addition, $H_{3D}(\bk)$ has two conserved spins $S_{X,Y}$ such that $[H_{3D}(\bk), S_{X,Y}] \PRLequal 0$.
%

\begin{figure}
\centering
\includegraphics[width=0.45\textwidth]{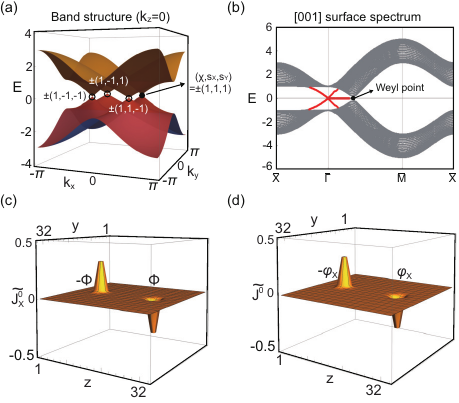}
\caption{
Tight-binding model $H_{3D}(\bk)$ for 3D semimetal with nonzero $Q_{iA}$.
(a) Energy spectra in the $k_z \PRLequal 0$ plane for $m \PRLequal-1$.
Eight Weyl points are located in pairs at $(\pm \pi/3, \pm \pi/3, 0)$.
The representative point $(\pi/3,\pi/3,0)$ is indicated by the filled black dots.
(b) [001] surface energy spectrum for a 20-layer slab.
The CSR modes (red color) are realized as the surface states of 3D semimetal, [cf. Fig.~\ref{fig1}(a)].
[(c) and (d)] A pair of (pseudospin) magnetic flux tubes with opposite signs is inserted along the translationally invariant $x$ direction, centered at $(y,z)=(16.5,9.5)$ and $(16.5,24.5)$, in a $32 \times 32$ system along the $y$ and $z$ directions.
(c) Pseudospin density $\mc{J}^0_X$ induced by magnetic flux $\Phi \PRLequal \pi/2$.
(d) Charge density $J^0$ induced by pseudospin magnetic flux $\vph_X \PRLequal \pi/2$.
Note that $\td{\mc{J}}^0_X \PRLequal \frac{4\pi^2}{Q_0} \mc{J}^0_X$ and $\td{J}^0 \PRLequal \frac{4\pi^2}{Q_0} J^0$.
}
\label{fig2}
\end{figure}

Let us focus on the regime $-2 \PRLless m \PRLless 0$, which is a semimetal where eight Weyl points are located at $\bb K \PRLequal (\pm K_*,\pm K_*,0)$ with $K_* \PRLequal \cos^{\PRLminus 1}(\PRLminus m/2)$ [see Fig.~\ref{fig2}(a)].
At each $\bb K$, there are two Weyl points with opposite chiralities, and they cannot be gapped because they are in different pseudospin sectors.
For example, the two Weyl points with chirality $\chi \PRLequal 1$ and $\PRLminus 1$ at $(K_*,K_*,0)$ have pseudospin eigenvalues $(1,1)$ and $(\PRLminus 1,\PRLminus 1)$, respectively [see Fig.~\ref{fig2}(a)].

With an open boundary along the $z$-direction, we can analytically derive the Hamiltonian $H_{\rm surf}(k_x,k_y)$ of the low-energy surface localized modes following Ref.~\onlinecite{dwivedi2016connecting}.
We find that $H_{\rm surf}(k_x,k_y) \PRLequal k_x S_Y \PRLminus k_y S_X$, which is exactly the CSR term, as detailed in Appendix~\ref{appendix:surf}.
The surface CSR modes exist only in a restricted region of the surface BZ, $\PRLminus K_* \le k_{x,y} \le K_*$, defined by the location of the Weyl points [Fig.~\ref{fig2}(b)].
Thus, the cutoff $\Lambda$ in the anomalous conservation laws for the CSR modes is naturally identified as $\Lambda=K_*$.
Below, we will study the bulk response of a 3D Weyl semimetal under external fields and see how the responses are related to the anomalous conservation laws of the 2D surface CSR modes derived in Eqs.~\eqref{eq:anomaly} and \eqref{eq:charge_anomaly_c4v} above.

\section{Bulk response and Weyl quadrupole}
Because of the anomalous response on the surface, we expect the Weyl semimetal to produce an interesting compensating bulk response.
In a simple time-reversal breaking Weyl semimetal having two Weyl nodes, there is an electromagnetic Chern-Simons response describing an anomalous quantum Hall effect.
Furthermore, the anomalous Hall coefficient is determined simply by the separation of two Weyl nodes in momentum space~\cite{zyuzin2012topological}, i.e., it is determined by the dipole moment of the Weyl nodes in momentum space.

In our time-reversal symmetric model Eq.~\eqref{eq:h3D}, the charge Chern-Simons response cancels out when the different pseudospin sectors are combined, and instead a mixed, spin-charge Chern-Simons response emerges:
\bg
S_{CS}
= \frac{e}{4\pi^2} \int d^4x \, \ep^{\mu \nu \rho \sg} Q_{\mu A} a^A_\nu \der_\rho A_\sigma.
\label{eq:CS_term}
\eg
Note that $\ep^{0123}=1$.
This term is {\it mixed} because it couples the charge $U(1)_{\rm EM}$ gauge field $A_\mu$ to the spin $U(1)_A$ gauge field $a_\mu^A$, and the tensor $Q_{\mu A}$ specifies the structure of this coupling.
The response coefficients $Q_{0A}=0$, and $Q_{iA}$ is the Weyl {\it spin-momentum} quadrupole moment defined as
\bg
Q_{iA}
= \sum_\alpha \chi_\alpha K_{\alpha,i} s_{\alpha,A},
\label{eq:smq}
\eg
where the $\alpha$ in Eq.~\eqref{eq:smq} is summed over all Weyl points sitting on the bulk FS.
The response in Eq.~\eqref{eq:CS_term} can be derived from the Kubo formula (See SM Sec.~S2~\cite{supple}).
Moreover, in Appendix~\ref{appendix:CS}, we provide an intuitive derivation based on the anomaly cancellation.
Namely, the currents induced by the bulk responses [Eqs.~\eqref{eq:sd} and \eqref{eq:cd}], at the open boundary, are canceled by the contribution from the surface CSR mode [Eq.~\eqref{eq:anomaly}].

Let us comment on the Weyl spin-momentum quadrupole moment $Q_{iA}$.
This matrix is well-defined when both the Weyl momentum, and spin, dipole moments vanish.
In our case, because of time-reversal symmetry the former vanishes, while a spin-resolved version of the Nielsen-Ninomiya no-go theorem~\cite{nielsen1981no} requires the latter to vanish.
Direct calculations of $Q_{iA}$ for our model Eq.~\eqref{eq:h3D} show that
$Q_{xY} \PRLequal Q_{yX} \PRLequal 0,$ and $Q_{xX} \PRLequal Q_{yY} \PRLequal 8 K_{*}\equiv Q_0$, i.e., $Q_{iA} = Q_0 \delta_{iA}$.
In fact, $Q_{iA}$ is proportional to the identity matrix $\delta_{iA}$ for $C_{nv}$ with $n \PRLgreater 2$, as detailed in SM Sec.~S2~\cite{supple}.

More explicitly, Eq.~\eqref{eq:CS_term} encodes the quasi-topological bulk charge and pseudospin responses to external fields;
a 3D conserved pseudospin Hall effect is given by $\mc{J}^i_A \PRLequal - \frac{e}{4\pi^2} Q_{kA} \ep^{ijk} E_j \PRLequal \sg_S \ep^{Aij} E_j$ where the pseudospin Hall conductivity $\sg_S \PRLequal -\frac{e}{4\pi^2} Q_0$.
Note that for a general symmetry class, the pseudospin Hall conductivity is $\sg^{ij}_A = - \frac{e}{4\pi^2} Q_{kA} \ep^{ijk}$.
Additionally, Eq.~\eqref{eq:CS_term} implies that  magnetic fields induce pseudospin density in the bulk via
\bg
\mc{J}^0_{X,3D}
= - \frac{e}{4\pi^2} Q_0 B_x, \,
\mc{J}^0_{Y,3D}
= - \frac{e}{4\pi^2} Q_0 B_y,
\label{eq:sd}
\eg
and, inversely, that the pseudospin magnetic fields bind electric charge density as
\bg
J^0_{3D} = - \frac{e}{4\pi^2} Q_0 (\mc{B}^X_x + \mc{B}^Y_y),
\label{eq:cd}
\eg
where $\mc{B}^A_i \PRLequal \epsilon_{ijk} \der_j a^A_k$ is the pseudospin magnetic field.

We confirm these bulk predictions with numerical calculations.
In a finite-size system with (32,32) unit cells along the $y$ and $z$ directions, we insert a pair of (pseudospin) magnetic flux tubes extended along the $x$ direction using Peierls substitution while maintaining translation symmetry along the $x$ direction.
Figures~\ref{fig2}(c) and \ref{fig2}(d) show the pseudospin (charge) density induced by the magnetic (pseudospin magnetic) field, localized around each flux tube, and the results match the predictions in Eqs.~\eqref{eq:sd} and \eqref{eq:cd}.
Details of the numerical calculation are provided in Appendix~\ref{appendix:numerics} and SM Sec.~S5~\cite{supple}.

Let us make a further remark about the bulk response.
We can generate a weak TI with $Q_{xX}=Q_{yY}=8\pi/a$ (where $a$ is the lattice constant) in our model Eq.~\eqref{eq:h3D} by annihilating the Weyl points having opposite chiralities.
This weak TI would exhibit a giant spin-Hall conductivity $\sigma_S=2e/(a\pi)$.
Since the weak TI relies on translation symmetry to be protected, we can introduce the translation-symmetry gauge field $\mathfrak{e}^{l}_{\mu}$ as discussed in Refs.~\onlinecite{nissinen2018tetrads,nissinen2019elasticity,gioia2021unquantized,chong2021emergent}, to find the effective action
\bg
S_{CS} = \frac{e}{4\pi^2} \int d^4x \, \ep^{\mu \nu \rho \sg} Q_{l A} \mf{e}^{l}_{\mu}  a^A_\nu \der_\rho A_\sigma.
\label{eq:CS_term3}
\eg
In addition to the responses above, this term indicates that chiral modes carrying pseudospins will appear at dislocations when the Burgers vector is in the $\hat{x}$ or $\hat{y}$ direction~\cite{ran2009one}.
Interested readers are referred to SM Sec.~S6~\cite{supple} for more details.
More remarks on the robustness of the bulk response under perturbations, and a connection to the symmetric rank-2 vector charge Chern-Simons term can be found in SM Secs.~S3 and S4~\cite{supple}.

\section{Conclusions}
Here we studied $C_{4v}$-symmetric SML terms with conserved spin, and discussed the dynamics and anomalies of a commuting spin Rashba term.
We showed that the CSR is anomalous by itself and can appear on the surface of a 3D topological semimetal.
Such a semimetal will have a vector charge Chern-Simons theory bulk response with a response coefficient determined by the Weyl spin-momentum quadrupole moment of the FS.
Beyond the specific crystalline symmetry group studied in this paper, our approach can be applied to general commuting SML terms in Eq.~\eqref{eq:surface_general}.
Interested readers can find relevant discussions in SM Sec.~S1~\cite{supple}.
This work opens an important new direction in the quasi-topological semimetal responses characterized by mixed multipole moments of the nodal FSes.
Furthermore, we expect this work to be relevant for both quantum materials and metamaterials.
Indeed, a precursor to this work~\cite{dubinkin2024higher}, where the vector charge was realized by momentum instead of spin, has recently been explored experimentally in a topo-electric circuit material~\cite{zhu2023higher}.
Also, the orbital degree of freedom used to construct pseudospins can be replaced with other internal degrees of freedom such as valley or sublattice.
These degrees of freedom can couple to strain fields and staggered potentials, which hence play the role of pseudospin gauge fields, and can be generated in, for example, graphene~\cite{pereira2009strain,levy2010strain,li2020valley} or cold atom systems~\cite{jaksch2003creation,aidelsburger2013realization}.
While our analysis focused on exactly conserved pseudospin, symmetry-allowed terms that weakly break pseudospin conservation can have different effects depending on the Weyl-node configuration: they can gap out coincident nodes, or, when the nodes are separated in momentum space, merely smooth the surface spectrum while leaving the Weyl nodes and their associated Fermi arcs intact.
Constructing an appropriate response theory in the presence of such pseudospin-breaking terms, in analogy with spin Chern numbers~\cite{prodan2009robustness} in systems without strict spin conservation or more general spin-resolved topology~\cite{kuansen2024spin}, is an interesting direction for future work.

\begin{acknowledgments}
Y.H. and P.Z. thank the US Office of Naval Research (ONR) Multidisciplinary University Research Initiative (MURI) Grant No. N00014-20-1-2325 on Robust Photonic Materials with High-Order Topological Protection.
Y.H. received additional support from the Air Force Office of Scientific Research under Award No. FA9550-21-1-0131.
T.L.H. thanks the ARO under the MURI Grant No. W911NF2020166 for support.
\end{acknowledgments}

\appendix

\section{Derivation of surface Hamiltonian $H_{\rm surf}(k_x,k_y)$}
\label{appendix:surf}
To derive the low-energy surface Hamiltonian $H_{\rm surf}(k_x,k_y)$ of the Weyl semimetal described by $H_{3D}(\bk)$ in Eq.~\eqref{eq:h3D}, let us expand Eq.~\eqref{eq:h3D} around the $\Gamma$ point in the BZ up to the first order in momentum $(k_x,k_y)$:
\ba
H_{3D}(\bk) \approx k_x \mu_x S_Y - k_y \mu_x S_X + k_z \mu_y + \delta m \mu_z,
\label{eq:h3D_approx}
\ea
where $\delta m = m+2$.
Note that we mainly focus on the regime $-2<m<0,$ and thus $\delta m > 0$ in the bulk.

We now consider an open boundary condition along the $z$-direction.
The boundary is located at $z=0$.
Below the boundary where $z<0$, we place the Weyl semimetal bulk.
This region is described by Eq.~\eqref{eq:h3D_approx} with $\delta m>0$.
The vacuum region is above the boundary, i.e. $z>0$.
Since the vacuum is adiabatically equivalent to a system with $\delta m<0$, we describe the vacuum region by Eq.~\eqref{eq:h3D_approx} with $\delta m<0$.
To solve this domain-wall problem, let us first focus on the case where $k_x = k_y = 0$, for simplicity.
By replacing $k_z$ by $-i \der_z$, we find four boundary modes that are exponentially localized along the $z$ direction.
These modes satisfy the equation:
\ba
(-i \der_z \mu_y + \delta m(z) \mu_z) \, \psi_*(z) = 0,
\ea
with solutions of the form
\ba
\psi_*(z) \propto e^{-\int^z_{-\infty} \, dz' \, \delta m(z')} \bpm 1 \\ 1 \epm_\mu \otimes v_s,
\label{eq:domain_solution}
\ea
where the column vector $(1,1)_\mu^T$ acts on the space of $\mu$ Pauli matrices, and
\bg
v_s \in \{(1,0,0,0)^T, (0,1,0,0)^T,(0,0,1,0)^T,(0,0,0,1)^T \}
\eg
corresponds to the four pseudospin sectors with $(S_X,S_Y)=(+,+), (+,-), (-,+), (-,-)$ in a diagonalized basis.

Then, by projecting the low-energy Hamiltonian in Eq.~\eqref{eq:h3D_approx} onto the Hilbert space spanned by the boundary mode $\psi_* (z)$, we immediately derive the surface Rashba term $H_{\rm surf}(k_x,k_y) = k_x S_Y -k_y S_X$ discussed in the main text.

\section{Chern-Simons response term $S_{CS}$ in Eq.~\eqref{eq:CS_term}}
\label{appendix:CS}
In this appendix, we will derive the spin-charge mixed Chern-Simons response term, which is shown in Eq.~\eqref{eq:CS_term} in the main text, in a heuristic way using the anomaly cancellation.
The main idea is as follows.
When an open boundary is made for our 3D Weyl semimetal described with Eq.~\eqref{eq:h3D}, the 2D surface CSR modes emerge with anomalous conservation laws [see Eqs.~\eqref{eq:anomaly} and \eqref{eq:charge_anomaly_c4v} in the main text].
Then, the bulk responses of a compensating Weyl semimetal must have a form such that their contribution to the open boundary (bulk inflow) must cancel the anomalous currents from the surface CSR modes.

Now consider a boundary at $z=0$ such that there is a nonvanishing (vanishing) bulk response in the system (vacuum) for $z<0$ ($z>0$).
Then we can model $Q_{iA}(z) \PRLequal Q^{(0)}_{iA} \Theta(\PRLminus z)$ as a $z$-dependent step function with $i=x,y,$ and $A=X,Y$ representing the spatial directions and pseudospin components respectively.
Recall that $Q^{(0)}_{iA} = Q_0 \delta_{iA}$ and $Q_0 = 8K_* = 8 \Lambda$ in our model $H_{3D}(\bk)$, where $K_*$ parameterizes the location of the Weyl points $\bb K = (\pm K_*, \pm K_*, 0)$ in the BZ, and $\Lambda$ is the momentum cutoff for computing the current responses of 2D CSR modes.

When the external spin electric fields $\mc{E}^X_y$ and $\mc{E}^Y_x$ are turned on, the surface 2D CSR modes exhibit the nonconserved electric currents on the boundary with
\bg
\der_\mu J^\mu_{\rm CSR}
= \frac{2e \Lambda}{\pi^2} (\mc{E}^X_y - \mc{E}^Y_x)
= \frac{e Q_0}{4\pi^2} (\mc{E}^X_y - \mc{E}^Y_x),
\eg
which is a rewriting of Eq.~\eqref{eq:charge_anomaly_c4v} in the main text.
To cancel the above contribution, the inflow of bulk current $J^\mu_{\rm Weyl}$ to the boundary must compensate $\der_\mu J^\mu_{\rm CSR}$, i.e. $\int dz \, \der_\mu J^\mu_{\rm Weyl} = - \der_\mu J^\mu_{\rm CSR}$.
Hence there will be an anomaly ascribed to the boundary:
\bg
\der_\mu J^\mu_{\rm Weyl}
\PRLequal \der_z J^z_{\rm Weyl}
\PRLequal \PRLminus \delta(z) \frac{Q_0 e}{4\pi^2} (\mc{E}^X_y \PRLminus \mc{E}^Y_x).
\label{eq:surf_current_app}
\eg
From this, we can infer that $J^z_{\rm Weyl} = \frac{Q_0 e}{4\pi^2} (\mc{E}^X_y \PRLminus \mc{E}^Y_x)$ in the Weyl semimetal region.
By recalling that the current is expressed by $J^\mu = \delta S_{eff} / \delta A_\mu$ with the effective action $S_{eff}$, we find $S_{eff} = \frac{Q_0 e}{4\pi^2} \int d^4 x \, A_z (\mc{E}^X_y - \mc{E}^Y_x) + \cdots$.
In a similar way, we can infer the effective action in terms of the spin electric fields with generic components, $\mc{E}^A_i = -\der_0 a^A_i + \der_i a^A_0$.
We can also repeat the same argument for the pseudospin current.
Thus, we find the mixed Chern-Simons term effective action,
\ba
S_{CS}
&= \frac{Q_0 e}{4\pi^2}\int d^4 x \, [ A_z (\mc{E}^X_y - \mc{E}^Y_x) + A_x \mc{E}^Y_z
\nn
& - A_y \mc{E}^X_z - A_0 (\mc{B}_x^X + \mc{B}_y^Y) ],
\label{eq:CS_term_app}
\ea
where $\mc{B}^A_i \PRLequal \epsilon_{ijk} \der_j a^A_k$ is the pseudospin magnetic field.

Furthermore, by noticing that $Q_{0A}=Q_{zA}=0$ and $Q_{iA}=Q_0 \delta_{iA}$ for $i=x,y$ in our model, we can express the Chern-Simons term by
\bg
S_{CS}
= \frac{e}{4\pi^2} \int d^4x \, \ep^{\mu \nu \rho \sg} Q_{\mu A} a^A_\nu \der_\rho A_\sigma,
\label{eq:CS_term_compact}
\eg
as in Eq.~\eqref{eq:CS_term}, for a general symmetry class.
Note that the Greek indices take on values $0,x,y,z,$ and $\ep^{0123}=\ep^{123}=1$.
The action $S_{CS}$ yields the bulk responses with
\bg
J^\mu_{3D} \PRLequal \PRLminus \frac{e}{4\pi^2 } \ep^{\mu i \nu \rho} Q_{iA} \der_\nu a^A_\rho,
\nn
\mc{J}^\mu_{A,3D} \PRLequal \PRLminus \frac{e}{4\pi^2} \ep^{\mu i \nu \rho} Q_{iA} \der_\nu A_\rho.
\label{eq:bulk_current}
\eg
Note also that without a boundary it is straightforward to see $\der_\mu J^\mu_{3D} \PRLequal \der_\mu \mc{J}^\mu_{A,3D} \PRLequal 0$, and hence they are conserved.

\begin{figure}[t!]
\centering
\includegraphics[width=1\columnwidth]{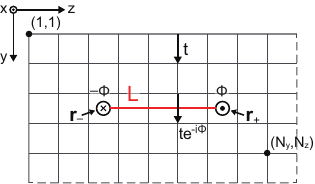}
\caption{
Pair of flux tubes.
Flux tubes with magnitudes $-\Phi$ and $\Phi$ are inserted at $(y,z)=\bb r_-$ and $\bb r_+$, respectively.
These flux tubes are stretched along $x$ direction.
The system is made with $(N_y,N_z)$ unit cells along the $y$ and $z$ directions where the periodic boundary conditions are imposed.
The red line $L$ connects two flux tubes.
The Peierls substitution is implemented such that a hopping amplitude $t$ is substituted by $t e^{\mp i\Phi}$ when it passes through $L$ depending on the direction of hopping process $\hat{t}$; $te^{-i\Phi}$ for $\hat{t} \cdot \hat{y}>0$ and $te^{i\Phi}$ for $\hat{t} \cdot \hat{y}<0$.
Pseudospin magnetic flux $\mc{B}^A_i$ can be implemented in a similar way.
For example, a pair of flux tubes for $\mc{B}^X_x$ can be described by the following Peierls substitution.
For hopping amplitudes pass through $L$, $t$ is substituted by $t e^{i S_X \vph_X}$ if $\hat{t} \cdot \hat{y}>0$ and $t e^{-i S_X \vph_X}$ if $\hat{t} \cdot \hat{y}<0$.
Here, $\vph_X$ is the flux of $\mc{B}^X_x$.
In our numerical calculation in Fig.~\ref{fig2}, $(N_y,N_z)=(32,32)$, $\bb r_- = (16.5,9.5)$, and $\bb r_+ = (16.5,24.5)$.}
\label{sfig1}
\end{figure}

\section{Numerical scheme for computing bulk responses in Eqs.~\eqref{eq:sd} and \eqref{eq:cd}}
\label{appendix:numerics}
In this appendix, we explain the numerical scheme for computing the bulk responses of our Weyl semimetal in the presence of external (pseudospin) magnetic fields (see also Fig.~\ref{fig2}).
The electromagnetic gauge field $A_i$ and the pseudospin gauge field $a_i^{A}$ enter the tight-binding model through Peierls substitution:
\ba
& h^{\alpha \beta}_{\bR, \bR'} \rightarrow h^{\alpha \beta}_{\bR, \bR'} e^{-i e \int_{\bR'+\bx_\beta}^{\bR + \bx_\alpha} \bb A \cdot d \bb l},
\\
& h^{\alpha \beta}_{\bR, \bR'} \rightarrow h^{\alpha \beta}_{\bR, \bR'} e^{i e \int_{\bR' + \bx_\beta}^{\bR + \bx_\alpha} \sum_A S_A^{\alpha \beta} \bb a^A \cdot d \bb l}.
\label{eq:peierls}
\ea
Here, $h^{\alpha\beta}_{\bR,\bR'}$ denotes the strength of hopping process of electron from $\beta$-th orbital within the unit cell $\bR'$ to the $\alpha$-th orbital within unit cell $\bR$.
(Thus $\bx_\alpha$ is the interunit cell position of the $\alpha$th orbital.)
As illustrated in Fig.~\ref{sfig1}, we insert a pair of magnetic (pseudospin magnetic) flux tubes along the $x$ direction in a periodic lattice by adding a phase $e^{-i\Phi}$ ($e^{i S_X \varphi_X}$) to the hoppings along  the $+y$ direction.

Then, the electric and pseudospin current densities $J^0_{3D}$ and $J^0_{A,3D}$ at each unit cell $\bb R$ are given by $\sum_{n \in {\rm occ}} {\rm Tr}_{\rm cell} [|\psi_n (\bb R)|^2]$ and $\sum_{n \in {\rm occ}} {\rm Tr}_{\rm cell} [S_A |\psi_n (\bb R)|^2]$, where $\psi_n(\bb R)$ is the wave function of the $n$th occupied eigenstate, and ${\rm Tr}_{\rm cell}$ is a trace over the internal degrees of freedom in each unit cell.
More details are provided in SM Sec.~S5~\cite{supple}.

\bibliography{refs.bib}

\end{document}


\title{Supplementary Material for ``Spin-momentum Locking and Topological Vector Charge Response with Conserved Spin"}

\author{Yoonseok Hwang}
\thanks{These two authors contributed equally.}
\affiliation{Department of Physics, University of Illinois Urbana-Champaign, Illinois 61801, USA}
\affiliation{Anthony J. Leggett Institute for Condensed Matter Theory, University of Illinois Urbana-Champaign, Illinois 61801, USA}

\author{Penghao Zhu}
\thanks{These two authors contributed equally.}
\affiliation{Department of Physics, University of Illinois Urbana-Champaign, Illinois 61801, USA}
\affiliation{Anthony J. Leggett Institute for Condensed Matter Theory, University of Illinois Urbana-Champaign, Illinois 61801, USA}

\author{Taylor L. Hughes}
\affiliation{Department of Physics, University of Illinois Urbana-Champaign, Illinois 61801, USA}
\affiliation{Anthony J. Leggett Institute for Condensed Matter Theory, University of Illinois Urbana-Champaign, Illinois 61801, USA}

\maketitle


\setcounter{section}{0}
\setcounter{figure}{0}
\setcounter{equation}{0}
\renewcommand{\thefigure}{S\arabic{figure}}
\renewcommand{\theequation}{S\arabic{equation}}
\renewcommand{\thesection}{S\arabic{section}}
\onecolumngrid
\tableofcontents
\hfill \\
\twocolumngrid

\section{$C_{n}$ and $C_{nv}$ symmetric spin-momentum locking terms in 2D}
\label{app:symmetry}
In this section, we study how the point groups in two dimensions (2D) constrain the spin-momentum locking term.
In 2D, point groups are classified into $C_n$ and $C_{nv}$ with $n=1,2,3,4,6$.
$C_{n}$ is the $n$-fold rotation symmetry around the axis perpendicular to the physical 2D plane, and the $C_{nv}$ symmetry group is generated by $C_{n}$ rotation and mirror symmetry along the $x$-direction, $M_{x}$.

\tocless{\subsection{General formula for leading terms}}{}
We start by considering the most general spin-momentum locking term in 2D which takes the form
\bg
\label{eq:sml}
d_x (k_x,k_y)S_X + d_y (k_x,k_y)S_Y + d_z (k_x,k_y) S_Z.
\eg
When time-reversal symmetry (TRS) $\mc{T}$ exists, $d_{x,y,z}$ must be odd in $k_{x,y}$.
In the presence of $C_{2,4,6}$ and $C_{2v,4v,6v}$ symmetries, $d_{z}$ should be even in $k_{x,y}$.
This implies that $d_z=0$ if we have both $C_{2,4,6}$ ($C_{2v,4v,6v}$) and TRS.
For $C_{3}$ and $C_{3v}$ symmetric systems, $d_z$ can be nonzero but the leading term is cubic in momentum: In the $C_{3}$ symmetric case, $d_{z}=k_{+}^{3}+k_{-}^{3}$ and/or $d_{z}=i(k_{+}^{3}-k_{-}^{3})$ where $k_{\pm}=k_{x}\pm i k_{y}$.
In the case of $C_{3v}$, because $M_{x}$ flips $S_Z$, $d_{z}$ can only be $k_{+}^{3}+k_{-}^{3}$ that is odd in $k_{x}$ \cite{fu2009hexagonal,samokhin2022effective}.
On the other hand, in $C_{1}$ and $C_{1v}$ symmetric systems, $d_{z}$ can be nonzero and linear in $k_{x,y}$.
In summary, for $C_{n}$ and $C_{nv}$ groups with $n \ge 2$, the most general spin-momentum locking terms that are linear in $k_{x,y}$ (i.e., at lowest order)~\cite{samokhin2022effective} take the form
\bg
\label{eq:g_hamiltonian}
h(\bk,\bb S) = \sum_{i,A} g_{iA} k_i S_A,
\eg
which is Eq.~(1) of the main text.
Here, $i=x,y$ and $A=X,Y$, $\bb S = (S_X,S_Y)$.
Under different symmetry groups, the allowed $g_{iA}$'s are different.
We will discuss the symmetry constraints for $g_{iA}$ in the next subsection.

\tocless{\subsection{Symmetry constraints of $g_{iA}$}}{}
\label{app:general_theory_sym}
Consider a symmetry operation $\sg$, which acts on real-space coordinate $\bb r$: $\bb r \to O_\sg \bb r + \bb \delta_\sg$.
Here, $O_\sg$ is an element in $O(3)$.
Hence, $O_\sg = l_\sg R_\sg$.
Note $l_\sg = {\rm Det} O_\sg \in \{+1,-1\}$.
$R_\sg$, which is $SO(3)$ part of $O_\sg$, can be characterized by $(\bb n, \phi)$, rotation axis and angle.
Thus, momentum and spin transform under $\sg$ as $\bk \to O_\sg \bk$ and $\bb S \to R_\sg \bb S$.
Thus, to be $h(\bk,\bb S) = h(O_\sg \bk, R_\sg \bb S)$, $g_{iA}$ must satisfy
\bg
\label{eq:g_symmetry}
g_{iA} = l_\sg [O_\sg]_{ij} g_{jB} [O_\sg^{-1}]_{BA}.
\eg

From Eq.~\eqref{eq:g_symmetry}, let us consider the constraints by $M_x$ and $C_4$.
Note that $M_x: (x,y) \to (-x,y)$ with $O_{M_x} = {\rm Diag}(-1,1)$ and $C_4: (x,y) \to (-y,x)$ with $O_{C_4} = \bpm 0 & -1 \\ 1 & 0\epm$.
Thus, $g_{iA}$ takes the form $g_{iA} = \bpm 0 & a_1 \\ a_2 & 0 \epm_{iA}$ under $M_x$ and $g_{iA} = \bpm a_1 & a_2 \\ -a_2 & a_1 \epm_{iA}$ under $C_4$.
More explicitly, $h(\bk,\bb S)$ are constrained by $M_x$ and $C_4$ as
\bg
M_x: a_1 k_x S_Y + a_2 k_y S_X \\
C_4: a_1 (k_x S_X + k_y S_Y) + a_2 (k_x S_Y - k_y S_X).
\eg
When $C_{4v} + \mc{T}$ exists, then
\bg
h(\bk,\bb S) = v (k_x S_Y - k_y S_X),
\eg
which was our primary focus in the main text.

\section{Bulk responses}
\label{app:kubo}

\tocless{\subsection{Derivation of linear responses}}{}
To study the mixed spin-charge linear response, we start from the Peierls substitution to derive the current operator. With charge and spin $U(1)$ gauge fields, the Hamiltonian can be expressed as
\ba
\label{eq:ps}
H =& \sum_{\mathbf{R}\mathbf{R}^{\prime}\alpha\beta} \, h_{\mathbf{R}\mathbf{R}^{\prime}}^{\alpha\beta} \, c_{\mathbf{R}\alpha}^{\dag} c_{\mathbf{R}^{\prime}\beta} \, e^{i 
 \phi_{\mathbf{R} \alpha, \mathbf{R'} \beta}},
\ea
where $\mathbf{R}, \mathbf{R}^{\prime}$ are coordinates of lattice sites, and $\mathbf{x}_\alpha$ denotes the position of $\alpha$-th orbital within unit cell.
The quantity $\phi_{\mathbf{R} \alpha, \mathbf{R'} \beta}$ is the Peierls phase,
\bg
\phi_{\mathbf{R} \alpha, \mathbf{R'} \beta}= \int_{\mathbf{R}^{\prime}+\mathbf{x}_\beta}^{\mathbf{R}+\mathbf{x}_\alpha} (-e\mathbf{A} + S^{\alpha\beta}_{A}\mathbf{a}^{A})\cdot d\mathbf{l}.
\eg
Note that we set $\hbar=c=1$ and the charge of electron to be $-e$. 

With spin $U(1)$ symmetries, $h_{\mathbf{R}\mathbf{R}^{\prime}}^{\alpha\beta}$ is nonzero only when $\alpha$ and $\beta$ are in the same spin sector.
With translation symmetry, $h^{\alpha\beta}_{\mathbf{R}, \mathbf{R}^{\prime}}$ depends on only the distance between two sites, i.e. $h^{\alpha\beta}_{\mathbf{R}\mathbf{R}^{\prime}}=h^{\alpha\beta}_{\mathbf{R}-\mathbf{R}^{\prime}}\equiv h^{\alpha\beta}_{\boldsymbol{\Delta \mathbf{R}}}$.
We Fourier transform and define
\bg
\label{eq:covention1}
c_{\mathbf{R} \alpha}^{\dag}
= \frac{1}{\sqrt{N}} \sum_{\mathbf{k}} e^{-i \bk \cdot (\mathbf{R} + \mathbf{x}_{\alpha})} c^{\dag}_{\bk \alpha}, 
\\
\label{eq:covention2}
h^{\alpha \beta}_{\bk}
= e^{-i \bf{k} \cdot (\mathbf{x}_{\alpha} - \mathbf{x}_{\beta})} \sum_{\Delta \mathbf{R}} e^{-i \bk \cdot \Delta \mathbf{R}} h^{\alpha \beta}_{\Delta \mathbf{R}},
\eg
where $N$ is the number of lattice sites.
Then, up to second order in gauge fields, Eq.~\eqref{eq:ps} can be written as
\ba
\label{eq:ps1}
H \simeq
& \frac{1}{N} \sum_{\Delta \mathbf{R} \mathbf{R}^{\prime} \alpha \beta} \sum_{\bk_{1} \bk_{2}} e^{-i\bk_{1} \cdot \Delta \mathbf{R}} e^{i(\bk_{2}-\bk_{1}) \cdot \mathbf{R}^{\prime}} h_{\Delta \mathbf{R}}^{\alpha \beta} 
 c^{\dag}_{\bk_{1} \alpha} c_{\bk_{2} \beta} \nn
& \times e^{-i \bk_1 \cdot \bx_\alpha + i \bk_2 \cdot \bx_\beta} \sum_{n=0}^{2} \frac{1}{n!} (i\phi_{(\mathbf{R}^\prime + \Delta \mathbf{R}) \alpha, \mathbf{R}^\prime \beta})^n
\\
&\equiv H_{0}+H_{\rm ext}.
\ea
We use an approximation
\bg
\phi_{\bR \alpha, \bR^\prime \beta}
\simeq \Delta \bb r \cdot (-e \bb A(\bb r) + S^{\alpha \beta}_A \bb a^A(\bb r))_{\bb r = \bb r_{\rm mid}}
\eg
where $\Delta \bb r = \bb r_1 - \bb r_2$ and $\bb r_{\rm mid} = \frac{1}{2}(\bb r_1 + \bb r_2)$ for $\bb r_1 = \bR + \bx_\alpha$, $\bb r_2 = \bR^\prime + \bx_\beta$.
We also define the Fourier transform of gauge fields,
\bg
\mathbf{A}(\mathbf{r})=\frac{1}{N}\sum_{\bk}e^{-i\bk \cdot\mathbf{r}}\mathbf{A}_\bk,
\\
\mathbf{a}^{A}(\mathbf{r})=\frac{1}{N}\sum_{\bk}e^{-i\bk \cdot\mathbf{r}}\mathbf{a}^{A}_\bk.
\eg
By substituting these equations in Eq.~\eqref{eq:ps1}, the term linear in gauge fields becomes
\bg
H_{\rm ext}^{(1)}
= \frac{i}{N} \sum_{\Delta \bR \bk \bb q} \sum_{\alpha\beta} e^{-i (\bk+\frac{\bb q}{2}) \cdot (\Delta \bR + \bx_\alpha - \bx_\beta)} h^{\alpha\beta}_{\Delta \bR} c_{\bk \alpha}^{\dag} c_{\bk+\bb q \beta} 
\nn
\times (\Delta \bR + \bx_\alpha - \bx_\beta) \cdot (-e\mathbf{A}_{\mathbf{q}} + S^{\alpha \beta}_{A} \mathbf{a}^{A}_{\mathbf{q}})
\nn
= - \frac{1}{N} \sum_{\bk \bb q} \sum_{\alpha \beta} c^{\dag}_{\bk-\frac{\bb q}{2} \alpha} c_{\bk+\frac{\bb q}{2} \beta} \frac{\der h^{\alpha \beta}_{\bk}}{\der \bk} \cdot (-e\mathbf{A}_{\bb q}+S^{\alpha \beta}_{A} \mathbf{a}^{A}_{\bb q}),
\label{eq:ext1}
\eg
and the term quadratic in gauge fields becomes
\bg
H_{\rm ext}^{(2)}
= \frac{1}{2N^2} \sum_{\bk \bb q \bb q^\prime} \sum_{\alpha\beta} c^{\dag}_{\bk-\frac{\bb q + \bb q^\prime}{2} \alpha} c_{\bk+\frac{\bb q + \bb q^\prime}{2} \beta} \frac{\der^2 h^{\alpha \beta}_{\bk}}{\der k_i \der k_j}
\nn
\times (-e \bb A_{\bb q} + S^{\alpha\beta}_{A} a^{A}_{\bb q})_i (-e \bb A_{\bb q^\prime} + S^{\alpha\beta}_{A} a^{A}_{\bb q^\prime})_j.
\label{eq:ext2}
\eg
We define the total charge and spin current at momentum $\mathbf{q}$ through
\bg
\label{eq:jk}
\mathbf{J}_{\text{tot}}(\mathbf{r})=\frac{1}{N}\sum_{\mathbf{q}}e^{-i\mathbf{q}\cdot\mathbf{r}}\mathbf{J}_{\text{tot}}(\mathbf{q}),
\\
\boldsymbol{\mathcal{J}}_{A,\text{tot}}(\mathbf{r})=\frac{1}{N}\sum_{\mathbf{q}}e^{-i\mathbf{q}\cdot\mathbf{r}}\boldsymbol{\mathcal{J}}_{A,\text{tot}}(\mathbf{q}).
\eg
The current operators, $\bb J_{\rm tot}(\bb q)$ and $\bb \mc{J}_{A, \rm tot}(\bb q)$, can be derived from $H_{\rm ext} = H^{(1)}_{\rm ext} + H^{(2)}_{\rm ext}$ through
\bg
\label{eq:jk1}
J^{i}_{\rm tot} (\bb q)
= -\frac{N}{V_{\rm cell}} \frac{\delta H_{\rm ext}}{\delta A_{i,-\bb q}},
\\
\label{eq:Jsk1}
\mc{J}^{i}_{A,\rm tot} (\bb q)
= -\frac{N}{V_{\rm cell}} \frac{\delta H_{\rm ext}}{\delta a^{A}_{i,-\bb q}},
\eg
where $V_{\rm cell}$ is the volume of the unit cell, and $\delta A_{i,\bb q}/\delta A_{j,\bb q^{\prime}} = \delta_{ij} \delta_{\bb q \bb q^{\prime}}$.
We now consider uniform gauge fields such that $A_{i,\bb q \ne \bb 0}=0$ and $a^A_{i,\bb q \ne \bb 0}=0$, and apply Eqs.~\eqref{eq:ext1}, \eqref{eq:ext2} and \eqref{eq:jk1}, we have
\bg
J^{i}_{\rm tot} = \frac{1}{V_{\rm cell}} \sum_{\bk \alpha \beta} c_{\bk \alpha}^{\dag} (J^{i}(\bk) + J^{i}_{\rm diam}(\bk))_{\alpha \beta} c_{\bk \beta},
\eg
where $J^{i}_{\rm tot}$ is short for $J^{i}_{\rm tot}(\bb q= \bb 0)$, and
\bg
J^{i}(\bk)=-e \der_{i} h_{\bk},
\\
J^{i}_{\rm diam}(\bk) = \frac{e}{N} \der_{i} \der_{j} h_{\bk} (-e A_{j}+S_{A}a^{A}_{j}),
\eg
are the paramagnetic current and diamagnetic current, respectively.
Note that $\der_{i}$ is short for $\der_{k_{i}}$, and $A_{i}$ ($a^{A}_{i}$) is short for $A_{i,\bb q= \bb 0}$ ($a_{i,\bb q = \bb 0}^{A}$).
Similarly, we have
\bg
\mc{J}^{i}_{A,\rm tot} = \frac{1}{V_{\rm cell}} \sum_{\bk \alpha \beta} c_{\bk \alpha}^{\dag}(\mc{J}^{i}_{A}(\bk) + \mc{J}^{i}_{A,\rm diam}(\bk))_{\alpha \beta} c_{\bk \beta},
\eg
where $\mc{J}^{i}_{A,\rm tot}$ is short for $\mc{J}^{i}_{A,\rm tot}(\bb q = \bb 0)$, and
\bg
\mc{J}^{i}_{A}(\bk) = S_{A} \der_{i} h_{\bk},
\\
\mc{J}^{i}_{A,\rm diam}(\bk) = -\frac{1}{N} S_{A} \der_{i} \der_{j} h_{\bk} (-e A_{j} + S_{A}a^{A}_{j}).
\eg
Then the conductivity for mutual charge-spin response at zero frequency is given by
\bg
\label{eq:conductivity}
\sg_{A}^{ij} = \lim_{\nu_{m} \rightarrow 0} \frac{\Pi^{ij}_{A}(i\nu_{m})}{\nu_{m}} + \sg_{A,\rm diam}^{ij},
\eg
where
\ba
\label{eq:kubo}
\Pi^{ij}_{A}(i\nu_m)
=& \frac{1}{\beta V} \sum_{\bk,n} \, {\rm Tr}[ \mc{J}^i_A(\bk) G(\bk,i\om_n+i\nu_m) \nn
& \times J^j(\bk) G(\bk,i\om_n)]
\ea
is the Kubo formula.
Here, $\beta=(k_{B}T)^{-1}$. $V=NV_{\rm cell}$ is the volume of the whole system. $\om_n = (2n+1)\pi/\beta$ is the Matsubara frequency for fermion.
Also, $G(\bk,i\om_n) = (i\om_n-h_{\bk})^{-1}$ is the Green function.
Meanwhile, the diamagnetic part $\sg_{A,\rm diam}^{ij}$ takes the form
\ba
\label{eq:cdiam}
&\sigma_{A,\rm diam}^{ij}=\lim_{\omega \rightarrow 0}\frac{1}{V}\sum_{\bk,a\in \rm occ} \frac{e}{i\omega} \langle a,\bk| S_{A} \der_i \der_j h_{\bk} |a,\bk\rangle
\nn
&=\lim_{\omega\rightarrow 0}\sum_{a\in \rm occ} \frac{e}{i\omega} \int \frac{d^d\bk}{(2\pi)^{d}} \langle a,\bk|S_{A} \der_{i} \der_{j} h_{\bk} |a,\bk\rangle,
\ea
which is divergent.
Importantly, we will show later that this divergence in $\sigma_{A,\rm diam}^{ij}$ cancels the divergence in the symmetric contribution to the off-diagonal $\sg_{xy}$ response in the Kubo formula [c.f. Eq.~\eqref{eq:kubo}].

To calculate the ${\rm Tr}[\dots]$ in Eq.~\eqref{eq:kubo}, we introduce energy eigenstates $\ket{a, \bk}$ such that $h_{\bk} \ket{a, \bk} = \ep_a(\bk) \ket{a, \bk}$.
Then,
\bg
G(\bk, i\om_n) = \sum_{a \in {\rm all}} \ket{a, \bk} \bra{a, \bk} \, [i\om_n-\ep_a(\bk)]^{-1} \\
{\rm Tr}[\dots]
= -e \sum_{a,b \in {\rm all}} \bra{a, \bk} S_A \der_i h_{\bk} \ket{b, \bk} \bra{b, \bk} \der_j h_{\bk} \ket{a, \bk} \nn
\times [i\om_n-\ep_a(\bk)]^{-1} [i\om_n+i\nu_m-\ep_b(\bk)]^{-1}
\eg
Note that 
\bg
\frac{1}{\beta} \sum_n \,  [i\om_n-\ep_a(\bk)]^{-1} [i\om_n+i\nu_m-\ep_b(\bk)]^{-1} \\
= \frac{\theta(E_F-\ep_a(\bk))-\theta(E_F-\ep_b(\bk))}{i\nu_m-\ep_b(\bk) + \ep_a(\bk)}
\eg
where the Fermi energy $E_F$.
Since $[h_{\bk},S_A]=0$ and $h_{\bk} \ket{a, \bk} = \epsilon_{a}(\bk) \ket{a, \bk}$, $\bra{a, \bk}S_A \der_i h_{\bk} \ket{b, \bk} = 0$ unless the spin eigenvalues of states $a$ and $b$ are the same, i.e. $S_{a,A} = S_{b,A}$.
We define $v_{iab}(\bk) = \bra{a, \bk} \der_i h_{\bk} \ket{b, \bk}$, and then $\Pi_{A}^{ij}(i\nu_{m})$ can be expressed as
\bg
\label{eq:corr}
\frac{e}{V}\sum_{\bk,a,\bar{b}}\left[\frac{v_{ja\bar{b}}(\bk)S_{a,A}v_{i\bar{b}a}(\bk)}{i\nu_{m}+\Delta_{\bar{b}a}(\bk)}-\frac{S_{a,A}v_{ia\bar{b}}(\bk)v_{j\bar{b}a}(\bk)}{i\nu_{m}-\Delta_{\bar{b}a}(\bk)}\right],
\eg
where $a$ labels eigenstates in the occupied subspace, $\bar{b}$ labels eigenstates in unoccupied subspace, and $\Delta_{\bar{b}a}(\bk)=\epsilon_{\bar{b}}(\bk)-\epsilon_{a}(\bk)$. The symmetric part of Eq.~\eqref{eq:corr} is then
\bg
\label{eq:corrsymm}
\frac{e}{V}\sum_{\bk,a,\bar{b}}\frac{\Delta_{\bar{b}a}(\bk)}{\nu_{m}^2+\Delta^2_{\bar{b}a}}S_{a,A}[v_{ia\bar{b}}(\bk)v_{j\bar{b}a}(\bk)+(i\leftrightarrow j)].
\eg
Replacing $\nu_{m}$ by $-i\omega$ and replacing the sum over $\bk$ by an integral, the symmetric part of $\Pi_{A}^{ij}$ contributes to the charge-spin mixed conductivity as
\ba
\lim_{\omega\rightarrow 0}\sum_{a,\bar{b}}&-\frac{e}{i\omega}\int \frac{d\bk}{(2\pi)^{d}}\Delta_{\bar{b}a}(\bk)S_{a,A}
\nn
\label{eq:conductivitysymm}
&\times[A_{ia\cm{b}}(\bk) A_{j\cm{b}a}(\bk) + (i\leftrightarrow j)].
\ea
where we define the non-Abelian Berry connection
\bg
A_{iab}(\bk) = -i \langle a, \bk| \der_i |b, \bk \rangle
\eg
and $v_{ia\bar{b}}(\bk) = i\Delta_{\bar{b}a}(\bk) A_{ia\cm{b}}(\bk)$ is used.
Next, we show explicitly that Eq.~\eqref{eq:cdiam} cancels Eq.~\eqref{eq:conductivitysymm}.
First, we use $\sum_{a\in \rm occ} \langle a,\bk| S_{A} \der_{i} \der_{j} h_{\bk} |a, \bk \rangle$ in Eq.~\eqref{eq:cdiam} is equal to
\bg
\sum_{a\in \rm occ} S_{a,A}[\der_{i} v_{jaa}(\bk) - \langle \der_{i} a,\bk| \der_{j} h_{\bk} |a, \bk \rangle
\nn
- \langle a, \bk |\der_{j} h_{\bk} |\der_{i} a,\bk \rangle],
\eg
where the first term vanishes after integrating over $\bk$ due to the periodicity.
Ignoring the first term, the above equation becomes
\ba
& i\sum_{a,a^{\prime}\in \rm occ} S_{a,A} [A_{iaa^\prime}(\bk) v_{ja^\prime a}(\bk) - v_{jaa^\prime}(\bk) A_{ia^\prime a}(\bk)]
\nn
& +i\sum_{a,\cm{b}} S_{a,A}[A_{ia\cm{b}}(\bk) v_{j\cm{b}a}(\bk) - v_{ja\cm{b}}(\bk) A_{i\cm{b}a}(\bk)],
\ea
where the first two lines vanish because $a$ and $a^{\prime}$ are symmetric.
Using $v_{ja\bar{b}}(\bk) = i\Delta_{\bar{b}a}(\bk) A_{ja\cm{b}}(\bk)$ again, we eventually have
\bg
\sum_{a\in \rm occ} \langle a, \bk| S_{A} \der_{i} \der_{j} h_{\bk} |a, \bk \rangle
\nn
= \sum_{a,\bar{b}} \Delta_{\bar{b}a}(\bk) S_{a,A} [A_{ia\cm{b}}(\bk) A_{j\cm{b}a}(\bk) + (i\leftrightarrow j)].
\eg
Substituting the above equation into Eq.~\eqref{eq:cdiam}, it is straightforward to see that $\sigma_{A,\rm diag}^{ij}$ cancels the symmetric part from the Kubo formula.

Therefore, only the antisymmetric part of the Kubo formula is important for the linear response, which takes the form
\bg
\Pi^{ij}_A(i\nu_m) = \frac{e}{V} \sum_{\bk,a,\cm{b}} \, \frac{i \nu_m}{[\ep_a(\bk) - \ep_{\cm{b}}(\bk)]^2} S_{a,A} \nn
\label{eq:Pi_1}
\times [v_{ia\cm{b}}(\bk) v_{j\cm{b}a}(\bk) - (i \leftrightarrow j)].
\eg
Note that the Abelian Berry connection and curvature are defined by
\bg
A_{i,a}(\bk) = A_{iaa}(\bk), \\
F^{a}_{ij}(\bk) = \der_i A_{j,a}(\bk) - \der_j A_{i,a}(\bk) \nn
= -i \sum_{\cm{b} \notin \rm occ} \left[A_{ia\cm{b}}(\bk) A_{j\cm{b}a}(\bk) - (i \leftrightarrow j)\right].
\eg
Then, Eq.~\eqref{eq:Pi_1} can be simplified to
\bg
\Pi^{ij}_A(i\nu_m) = -\frac{e}{V} \nu_m \sum_{\bk,a \in \rm occ} S_{a,A} F_{ij}^{a}(\bk).
\eg
Thus, the conductivity $\sg_A^{ij}$ becomes
\ba
\sg^{ij}_A
=& \lim_{\nu_m \to 0} \frac{1}{\nu_m} \Pi^{ij}_A(\bk) \\
=& - \frac{e}{(2\pi)^3} \sum_{a\in \text{occ}} \int d\bk S_{a,A} F_{ij}^{a}(\bk).
\ea
From the Stokes theorem,
\bg
\int d\bk \sum_{a\in \text{occ}} S_{a,A}F_{ij}^{a}(\bk)
\nn
=\sum_{\gamma}\int_{FS_{\gamma}} \ep_{ijk} k_k \sum_{a\in\text{occ}} S_{a,A} \left( \frac{1}{2} F_{\mu\nu}^{a}(\bk) dS_\mu \wedge dS_\nu \right)
\nn
= 2\pi\epsilon_{ijk} \sum_\alpha \chi_\alpha K_{\alpha,k} S_{\alpha, A}
= 2\pi\epsilon_{ijk} Q_{kA}.
\eg
Here $FS_\gamma$ denotes a set of disjoint Fermi surfaces.
The index $\alpha=(\gamma,a)$ in the last line labels both the pseudospin sector and the location in momentum space of the Fermi surfaces. 
Note that we set the Fermi level slightly above/below the Weyl points. Hence, we arrive at
\bg
\label{eq:conductivity_tensor}
\sg^{ij}_A = -\frac{e}{4\pi^2} \ep_{ijk} Q_{kA}.
\eg
Then, from $J^i_{\rm bulk} = \sg^{ij}_A \mc{E}^A_j$ and $\mc{J}^i_A = \sg^{ij}_A E_j$, it is straightforward to derive the bulk responses shown in the main text.

\tocless{\subsection{Symmetry constraints on $Q_{iA}$}}{}
In the main text, we showed that the Weyl spin-momentum quadrupole $Q_{iA}$ plays the role of a response coefficient.
In this subsection we discuss the symmetry constraints on $Q_{iA}$.
We first note that the chiral charge $\chi_\alpha$, momentum $\bb K_\alpha$, and pseudospin $\bb S_\alpha$ of Weyl point $\alpha$ transform like a pseudoscalar, a vector, and an axial vector, respectively.
Thus, for a symmetry $\sg$ with $O(3)$ element $O_\sg$ (see Sec.~\ref{app:symmetry} for the definition of $O_\sg$), we derive
\bg
\label{eq:Q_symmetry}
Q_{iA} = [O_\sg]_{ij} Q_{jB} [O_\sg^{-1}]_{BA}.
\eg

For example, mirror $M_x$ and $C_n$ rotation impose the following constraints on $Q_{iA}$:
\bg
M_x: Q_{iA}=\bpm a_1 & 0 \\ 0 & a_2 \epm_{iA}, \\
C_2: Q_{iA}=\bpm a_1 & a_2 \\ a_3 & a_4 \epm_{iA}, \\
C_{n>2}: Q_{iA}=\bpm a_1 & a_2 \\ -a_2 & a_1 \epm_{iA},
\eg
where $a_{1,2,3,4} \in \mathbb{R}$.
This implies that $Q_{iA}$ is proportional to the identity matrix when $C_{nv}$ ($n>2$) exists, in a suitable coordinate system.
\\

\section{Perturbations effect on the commuting-spin Rashba mode and bulk responses}
\label{app:symbrk}
In this section, we discuss the symmetric perturbation terms allowed in $C_{4v}+\mc{T}$ symmetric class.
Up to the quadratic order in $\bk$, there exist only two perturbation terms that preserve both $C_{4v}+\mc{T}$ and spin-$U(1)$ symmetries, and they can be added into the commuting-spin Rashba (CSR) Hamiltonian:
\ba
h_\bk = & v k_{x}S_{Y}- 
v k_{y}S_{X}
\nn
\label{eq:perturb}
& + \alpha (k_{x}^2+k_{y}^2) \mathds{1}_{4 \times 4} + \beta k_{x} k_{y} S_{X} S_{Y}.
\ea
As mentioned in the main text, these two perturbation terms can close the Fermi surface and make 2D CSR mode not anomalous:
$h_\bk$ has four energy bands $E_{\bk}^{1,2} = \alpha (k_{x}^2+k_{y}^2) \pm v(k_{x}-k_{y}) + \beta k_{x}k_{y}$ and $E_{\bk}^{3,4} = \alpha (k_{x}^2+k_{y}^2)\pm v(k_{x}+k_{y}) - \beta k_{x}k_{y}$.
The Fermi surfaces of these four bands are generally conic sections that are closed when $|\alpha|>|\beta|/2$ if $\bk$ can be arbitrarily large.
More precisely, in order for theses Fermi surfaces to be closed, the momentum cutoff $\Lambda$ should be larger than a bound, $2(|v(\alpha+\beta/2)| + \sqrt{v^2(\alpha+\beta/2)^2 + (\alpha^2-\beta^2/4)(v^2+4\mu\alpha)})/(4\alpha^2-\beta^2)$, where $\mu$ is the chemical potential.
%

\begin{figure}
\centering
\includegraphics[width=1\columnwidth]{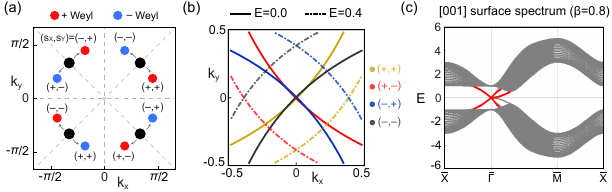}
\caption{
{\bf Effect of symmetry-preserving perturbation.}
(a) Splitting of Weyl points in the $k_z=0$ plane by the perturbation $\beta \sin k_x \sin k_y \mu_x S_X S_Y$.
Before adding the perturbation ($\beta=0$), two Weyl points are degenerate at the gray dots.
(b) Fermi-energy contours for $\beta=0.8, m=-1$.
(c) [001] surface energy spectrum for 20 layers when $\beta=0.8$ and $m=-1$.
The deformed CSR surface modes on the top surface are indicated by the red color.
}
\label{sfig1}
\end{figure}

For the surface CSR mode of the 3D topological semimetal [c.f. Eq.~(10) in the main text], the two allowed perturbation terms can be added through bulk terms $(2-\cos k_{x}-\cos k_{y})\mu_{i}\mathbbm{1}_{4\times 4}$ and $\sin k_{x}\sin k_{y}\mu_{i}S_{X}S_{Y}$, where $i=0,x,y,z$ and $\mu_{0}$ is the $2\times 2$ identity matrix.
The terms with $\mu_{0}$ only shift the energy of the Weyl points, and the terms with $\mu_{x,y,z}$ can move and/or split the Weyl points in the momentum space.
We show the splitting of the Weyl point pairs, and the deformation of the surface mode, induced by a nontrivial symmetry preserving perturbation term, $\beta \sin k_x \sin k_y \mu_x S_X S_Y$ in Fig.~\ref{sfig1}. 
However, this deformation does not qualitatively affect the bulk response and the anomalies of the surface mode, which are still determined by $Q_{iA}$.
\\

\section{Anomalies of the generalized spin-momentum locked mode}
\label{app:general_theory}
\tocless{\subsection{Current anomaly}}{}
\label{app:general_theory_anomaly}
In this section, we explicitly write down the anomalous conservation laws for the generalized Hamiltonian in Eq.~\eqref{eq:g_hamiltonian}.
To simplify the equations, we define $g_{i\pm} = g_{iX} \pm g_{iY}$, and $\sg_{i\pm} = \sgn (g_{i\pm})$.
The anomalous conservation laws for charge and spin currents can be obtained by inserting gauge fluxes in the same way described in the main text.
Then, the divergence of charge and spin currents are given by
\bg
\der_\mu \mc{J}^\mu_X
= -\frac{e\Lambda_y}{\pi^2} (\sg_{x+} + \sg_{x-}) E_x - \frac{e\Lambda_x}{\pi^2} (\sg_{y+} + \sg_{y-}) E_y,
\nn
\der_\mu \mc{J}^\mu_Y
= -\frac{e\Lambda_y}{\pi^2} (\sg_{x+} - \sg_{x-}) E_x - \frac{e\Lambda_x}{\pi^2} (\sg_{y+} - \sg_{y-}) E_y,
\nn
\der_\mu J^\mu
= - \frac{e\Lambda_x}{\pi^2} \large[ \sg_{y+} (\mc{E}^X_y+\mc{E}^Y_y) + \sg_{y-} (\mc{E}^X_y-\mc{E}^Y_y) \large] \nn - \frac{e\Lambda_y}{\pi^2} \large[ \sg_{x+} (\mc{E}^X_x+\mc{E}^Y_x) + \sg_{x-} (\mc{E}^X_x-\mc{E}^Y_x) \large].
\label{eq:response_all}
\eg
The last equation can be further simplified if we introduce $\mc{E}^{\pm}_i = \mc{E}^X_i \pm \mc{E}^Y_i$:
\bg
\der_\mu J^\mu
= - \frac{e\Lambda_x}{\pi^2} (\sg_{y+} \mc{E}^+_y + \sg_{y-} \mc{E}^-_y) \nn
- \frac{e\Lambda_y}{\pi^2} (\sg_{x+} \mc{E}^+_x + \sg_{x-} \mc{E}^-_x).
\eg

This reproduce the anomalous conservation laws discussed in the main text.
By substituting $g_{xY}=1$, $g_{yX}=-1$, and $g_{xX}=g_{yY}=0$,
\bg
\der_\mu \mc{J}^\mu_X = \frac{2e \Lambda_x}{\pi^2} E_y, \quad
\der_\mu \mc{J}^\mu_Y = - \frac{2e \Lambda_y}{\pi^2} E_x, \nn
\der_\mu J^\mu = \frac{2e \Lambda_x}{\pi^2} \mc{E}^X_y - \frac{2e \Lambda_y}{\pi^2} \mc{E}^Y_x.
\eg
Note that $\Lambda_x = \Lambda_y = \Lambda$ is set in the main text because of $C_{4v}$ symmetry.

\tocless{\subsection{Anomaly cancellation and spin-momentum quadrupole}}{}
\label{app:quadrupole}
Generalized spin-momentum locked modes cannot exist solely because of the current anomalies.
As discussed in the main text, they can live on the surface of 3D semimetals with proper spin-momentum quadrupole $Q_{iA}$.
We now discuss what the word \emph{proper} means in terms of the Callan-Harvey anomaly cancellation mechanism~\cite{callan1985anomalies}.
The bulk response term determined by the spin-momentum quadrupole generally takes the form
\bg
S_{CS}[A,a]
= \int d^4x \alpha \ep^{\mu\nu\rho\sg} Q_{\mu A} \der_\nu a^A_\rho A_\sg, \\
J^\mu
= \frac{\delta S}{\delta A_\mu}
= - \alpha \ep^{\mu i \nu \rho} Q_{iA} \der_\nu a^A_\rho, \\
\mc{J}^\mu_A
= \frac{\delta S}{\delta a^A_\mu}
= - \alpha \ep^{\mu i \nu \rho} Q_{iA} \der_\nu A_\rho.
\eg
Note that the Greek indices take on values $0,x,y,z$ and $\ep^{0123}=-\ep_{0123}=\ep^{123}=1$.
When the bulk has a constant spin-momentum quadrupole that is denoted as $Q_{0,iA}$, then the presence of boundary with $Q_{iA}(z) = Q_{0,iA} \theta(z)$,
this contributes $\der_\mu J_b^\mu = \int_{0_-}^{0^+} dz \der_\mu J^\mu$ and $\der_\mu \mc{J}_{b,A}^\mu = \int_{0_-}^{0^+} dz \der_\mu \mc{J}_{b,A}^\mu$.
Hence, considering the contributions from both surface and bulk, we have
\bg
\der_\mu (J_b^\mu + J_s^\mu) = -f_{iA} \mc{E}^A_i, \\
\der_\mu (J_{b,A}^\mu + J_{s,A}^\mu) = -f_{iA} E_i,
\eg
where the coefficients $f_{iA}$ are given by
\bg
f_{xX} = \alpha Q_{0,yX} +\frac{e\Lambda_y}{\pi^2} (\sg_{x+}+\sg_{x-}), \\
f_{xY} = \alpha Q_{0,yY} +\frac{e\Lambda_y}{\pi^2} (\sg_{x+}-\sg_{x-}), \\
f_{yX} = -\alpha Q_{0,xX} +\frac{e\Lambda_x}{\pi^2} (\sg_{y+}+\sg_{y-}), \\
f_{yY} = -\alpha Q_{0,xY} +\frac{e\Lambda_x}{\pi^2} (\sg_{y+}-\sg_{y-}).
\eg
All of the four $f_{iA}$ must be zero to have the anomaly cancellation between bulk and surface.
This is equivalent to say that generalized spin-momentum locked modes can only live on the surface of  3D systems with bulk spin-momentum quadrupole satisfying $f_{iA}=0$.

Finally, we comment that, in the presence of symmetries, $Q_{iA}$ and $g_{iA}$ are further constrained by Eqs.~\eqref{eq:g_symmetry} and \eqref{eq:Q_symmetry}.
\\

\tocless{\subsection{Example: $C_{4v}$ symmetry class with the alternative $C_{4z}$}}{}
\label{app:c4z_symmetric}
In the main text and Sec.~\ref{app:symmetry} and the last subsections, we derive the surface CSR Hamiltonian and bulk response in $C_{4v}+\mc{T}$ symmetric class.
Here, we choose $C_{4z}$ rotation such that $C_{4z}:(x,y,z) \to (- y,x,z)$ and $(S_{X},S_{Y}) \to (S_{Y},-S_{X})$.
The, the anomalous charge conservation law is given by
\bg
\der_\mu J^\mu = \frac{2e \Lambda}{\pi^2}(\mc{E}^X_y - \mc{E}^Y_x).
\eg

However, one can also consider an alternative fourfold rotation symmetry $\td{C}_{4z}$ rotation with $\td{C}_{4z}:(x,y,z) \to (-y,x,z)$ and $(S_{X},S_{Y}) \to (- S_{Y},S_{X})$.
Now, the pseudospin rotates opposite to our previous convention.
According to Eqs.~\eqref{eq:g_symmetry} and \eqref{eq:response_all}, this change leads to a CSR term $h_{\bk} = v(k_{x}S_{Y} + k_{y}S_{X})$ and the charge anomaly with
\bg
\der_\mu J^\mu = \frac{2e \Lambda}{\pi^2}(\mc{E}^X_y + \mc{E}^Y_x).
\eg
Note that now $\der_\mu J^\mu$ is induced by the symmetric combination of pseudospin electric fields, i.e., $\mc{E}^X_y + \mc{E}^Y_x$.

Charge and pseudospin responses with the alternative $\td{C}_{4z}$ can be obtained by a transformation $(\mc{E}^{Y}_{i}, \mc{B}^{Y}_{i}) \rightarrow -(\mc{E}^{Y}_{i}, \mc{B}^{Y}_{i})$ from the responses with $C_{4z}$.
With this change, the bulk response becomes a pseudospin version of the symmetric rank-2 vector charge Chern-Simons term discussed in Refs.~\onlinecite{dubinkin2024higher,zhu2023higher}, which is the bulk response term for a 3D system with surface rank-2 chiral mode.
\\

\section{Numerical scheme for linear response}
\label{app:numerics}
In the main text, we studied the bulk responses under the electromagnetic (EM) $U(1)_{\rm EM}$ and pseudospin $U(1)_{A=X,Y}$ magnetic fields.
Here, we explain how (pseudospin) magnetic fluxes are implemented in our tight-binding model.
%

\begin{figure}
\centering
\includegraphics[width=0.8\columnwidth]{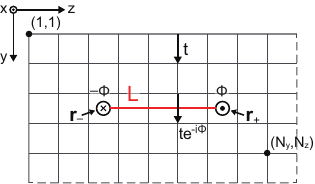}
\caption{
{\bf A pair of flux tubes.}
Flux tubes with magnitudes $-\Phi$ and $\Phi$ are inserted at $(y,z)=\bb r_-$ and $\bb r_+$, respectively.
These flux tubes are stretched along $x$ direction.
The red line $L$ connects two flux tubes.
The Peierls substitution is implemented such that a hopping amplitude $t$ is substituted by $t e^{\mp i\Phi}$ when it passes through $L$ depending on the direction of hopping process $\hat{t}$;
$te^{-i\Phi}$ for $\hat{t} \cdot \hat{y}>0$ and $te^{i\Phi}$ for $\hat{t} \cdot \hat{y}<0$.
Pseudospin magnetic flux $\mc{B}^A_i$ can be implemented in a similar way.
For example, a pair of flux tubes for $\mc{B}^X_x$ can be described by the following Peierls substitution.
For hopping amplitudes pass through $L$, $t$ is substituted by $t e^{i S_X \vph_X}$ if $\hat{t} \cdot \hat{y}>0$ and $t e^{-i S_X \vph_X}$ if $\hat{t} \cdot \hat{y}<0$.
Here, $\vph_X$ is the flux of $\mc{B}^X_x$.
}
\label{sfig2}
\end{figure}

First, let us explain the case of $U(1)_{\rm EM}$ magnetic field.
The magnetic field can be defined by spatial components of gauge field $A_\mu$.
In the presence of $A_\mu$, the hopping amplitudes $h^{\alpha\beta}_{\bR,\bR^\prime}$ are changed by
\bg
h^{\alpha\beta}_{\bR,\bR^\prime} e^{-i e \int_{\mathbf{R}^{\prime}+\mathbf{x}_\beta}^{\mathbf{R}+\mathbf{x}_\alpha} \mathbf{A}\cdot d\mathbf{l}}
\eg
according to the Peierls substitution.
Note that we use the same tight-binding notation introduced in Sec.~\ref{app:kubo}.

Let us consider the magnetic field $B_x$ along the $x$ direction.
For this, we introduce a pair of flux tubes with magnitudes $\Phi$ and $-\Phi$.
In this case, we can impose the periodic boundary condition along all three directions.
Moreover, $k_x$ remains as well-defined quantum number because the flux tubes keep the translation symmetry along the $x$ direction.
A schematic description of flux tubes is shown in Fig.~\ref{sfig2}.
The configuration of gauge field $A_\mu$, which is compatible with the flux tubes at $(y,z)=\bb r_-$ and $\bb r_+$, is described by a string connecting two flux tubes.
(In fact, it is a strip not string because flux tubes are stretched along the $x$ direction.)
For a given string $L$, the Peierls substitution can be easily implemented.
When a hopping amplitude $t$ passes through $L$, then it is substituted by $te^{\mp i\Phi}$ depending on the direction of hopping process (see Fig.~\ref{sfig2} for the details).
Note that we normalize the magnetic flux $\Phi$ such that $\Phi=2\pi$ is equivalent to a magnetic flux quantum $\phi_0=2\pi/e$.
Thus, $\Phi = 2\pi \phi/\phi_0$ where $\phi$ is the magnetic flux per unit cell.
The same normalization is also applied to the pseudospin magnetic flux.

The pseudospin magnetic flux can be implemented in a similar way.
For the pseudospin gauge field $a^A_i$, the relevant Peierls substitution is defined by
\bg
h^{\alpha\beta}_{\bR,\bR^\prime} e^{i e \int_{\bR^{\prime} + \bx_\beta}^{\bR+\bx_\alpha} \sum_A S_A^{\alpha\beta} \bb a^A \cdot d\mathbf{l}}.
\eg
Thus, for a hopping amplitude $t$ passes through $L$, the Peierls substitution leads $t \to te^{\pm i S_A \vph_A}$, where $\vph_A$ is the spin magnetic flux corresponding to $U(1)_A$ symmetry.

As representative cases, we considered (1) a pair of magnetic flux tubes with strength $\Phi=\pm \pi/2$ and (2) a pair of pseudospin magnetic flux tubes with strength $\vph=\pm \pi/2$, in the model Hamiltonian in Eq.~(10) presented the main text.
Note that flux tubes stretches along the $x$ direction and are located at $(y,z)=(N_y/2+1/2,N_z/4+1/2)$ and $(N_y/2+1/2,3N_z/4+1/2)$ in both calculations.
Here, $N_y=32$ and $N_z=32$ are the dimension of the lattice along $y$ and $z$ directions.
To compute the charge and spin densities, $J^0(y,z)$ and $\mc{J}^0_A(y,z)$, we take a half filling and sum the square of eigenstates weighted by charge and spin over filled states.
Then, these values are averaged for $N_k=21$ different $k_x$ values, which corresponds to $N_x=21$ system size along the $x$ direction.
[See Fig.~2 in the main text for the distribution of $J^0(y,z)$ and $\mc{J}^0_A(y,z)$.]

Our theoretical analysis predicts $\int_{\rm flux} dy dz \mc{J}^0_X(y,z) = -\frac{Q_0}{4\pi^2} \Phi$ in the case (1), and $\int_{\rm flux} dy dz J^0(y,z) = -\frac{Q_0}{4\pi^2} \vph$ in the case (2), respectively.
Here, $Q_0=Q_{xX}=8\pi/3$ is the nonzero component in $Q_{iA}$, and $\int_{\rm flux}$ means that the charge/pseudospin density is summed over a restricted region close to one of flux tubes.
When 16 sites near a flux tube is considered, the numerical result is consistent with the predicted value within 8.08\% error.
The deviation of numerical result from the theoretical prediction becomes reduced if more sites are included.
For example, the deviation is 4.01\% if 64 sites considered.

\section{Weak TI and response to dislocations}
\label{app:retodislocation}
In this section, we discuss the response to dislocations of the weak TI  with nonzero  pseudospin-momentum quadrupole $Q_{iA}$. 
%

\begin{figure}
\centering
\includegraphics[width=1\columnwidth]{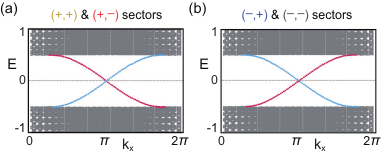}
\caption{
{\bf Chiral modes localized on the screw dislocations.}
Energy spectrum of pseudospin sectors (a) $(+,+)$ and $(+,-)$ and (b) $(-,+)$ and $(-,-)$ of the weak TI model $H_{3D}(\bk,t=1)$.
Note that in our numerical calculation, we have a lattice that has $24 \times 24$ sites in the $yz$-plane, and has two screw dislocations with Burgers vectors $\mathbf{b}=\pm a \mathbf{e}_{x}$ along the $x$-direction.
The center of the screw dislocations are located at $(y,z)=(12,12)$ and $(12,24)$.
Modes highlighted by red
(blue) color are chiral modes localized on the dislocation with
Burgers vector $b = a\mathbf{e}_{x}$ ($b = -a\mathbf{e}_{x}$).
}
\label{sfig3}
\end{figure}

First, we consider a continuous deformation from the Weyl semimetal, which is described by Eq.~(10) in the main text, to a weak TI.
To this end, we define a one-parameter family of Hamiltonian,
\bg
\label{eq: weakTI}
H_{3D}(\bk,t) = (1-t)(\sin k_x \mu_x S_Y -\sin k_y \mu_x S_X) \nn
+ [m -1 + (1-t)(\cos k_x + \cos k_y) + \cos k_z] \mu_z \nn
+ \sin k_z \mu_y + t \mu_{x}
U_{1} + t \mu_{z} U_{2},
\eg
which connects the Weyl semimetallic phase and weak TI phase adiabatically.
The diagonal matrices $U_{1}$ and $U_{2}$ takes the form $U_1={\rm Diag} (\sin(k_{x}-k_{y}), \, \sin(-k_{x}+k_{y}), \, \sin(-k_{x}-k_{y}), \, \sin(k_{x}+k_{y}))$ and $U_2={\rm Diag} (\cos(k_{x}-k_{y}), \cos(-k_{x}+k_{y}), \cos(-k_{x}-k_{y}), \cos(k_{x}+k_{y}))$ under the basis where $S_X={\rm Diag} (1,-1,1,-1)$ and $S_Y={\rm Diag} (1,-1,-1,1)$.
Also, we set $m=-0.5$ throughout this section.
By tuning $t$ from $0$ to $1$, one can continuously move the Weyl points with opposite chiralities toward $\bk=(\pi,\pi,0)$, and then gap them out to derive a weak TI.

Then, let us consider screw dislocations along the $x$-direction in the weak TI model $H_{3D}(\bk,t=1)$.
We require periodic boundary conditions along all three directions, and thus must have two screw dislocations that have opposite Burgers vectors $\mathbf{b}=\pm a\mathbf{e}_{x}$, where $a$ and $\mathbf{e}_{x}$ are the lattice constant and unit vector along the $x$-direction respectively.
From numerical calculations, we observe chiral and anti-chiral modes trapped on the dislocations in different pseudospin sectors, which is consistent with the cut and glue picture discussed in Ref.~\onlinecite{ran2009one}.
As shown in Fig.~\ref{sfig3}(a), in pseudospin sector $(+1,+1)$ and $(+1,-1)$, there is a chiral mode with positive (negative) chirality on the screw dislocation with Burgers vector $\mathbf{b}=-a\mathbf{e}_{x}$ ($\mathbf{b}=a\mathbf{e}_{x}$).
In contrast, as shown in Fig.~\ref{sfig3}(b), in pseudospin sector $(-1,+1)$ and $(-1,-1)$, there is a chiral mode with negative (positive) chirality on the screw dislocation with Burgers vector $\mathbf{b}=-a\mathbf{e}_{x}$ ($\mathbf{b}=a\mathbf{e}_{x}$).
Therefore, a single screw dislocation does not produce a charge anomaly, but can produce a pseudospin anomaly, which is captured by the action,
\ba
S_{CS} &= \frac{e}{4\pi^2} \int d^4x \ep^{\mu \nu \rho \sg} Q_{l A} \mathfrak{e}^{l}_{\mu}  a^A_\nu \der_\rho A_\sigma.
\ea
By variations with respect to charge or pseudospin gauge fields, we can derive the charge or pseudospin current as
\bg
J^\mu=\frac{e}{4\pi^2} \ep^{\mu\nu\rho\sg} Q_{lA} \der_\rho (\mf{e}^l_\sg a^A_\nu),
\\
\mc{J}^\mu_A = - \frac{e}{4\pi^2} \ep^{\mu\nu\rho\sg} Q_{lA} \mf{e}^l_\nu \der_\rho A_\sg.
\eg
It is straightforward to observe that $J^{\mu}$ is conserved and satisfies $\partial_{\mu}J^\mu=0$.
However, $\mc{J}^\mu_A$ has an anomalous conservation law:
\bg
\label{eq:acSM}
\partial_{\mu}\mc{J}^\mu_A = - \frac{e}{4\pi^2} \ep^{\mu\nu\rho\sg} Q_{lA} \partial_{\mu}\mf{e}^l_\nu \der_\rho A_\sg,
\eg
which says that if we have both nonzero torsional magnetic (electric) field and electric (magnetic) field, then the pseudospin current is anomalous.
In lattice systems, a dislocation can provide nonzero torsional magnetic flux (i.e., the flux of translation gauge field $\mf{e}^l_\nu$).
For example, a screw dislocation with Burgers vector $\mathbf{b}=-a \mathbf{e}_{x}$ will lead to a nonzero torsional flux $\int dydz (\partial_{y}\mf{e}^x_{z}-\partial_{z}\mf{e}^x_{y})=-a$ localized on the screw dislocation.
With such a dislocation, if we furthermore have a nonzero electric field along the $x$-direction, then the pseudospin current will be anomalous and satisfies $\int dy dz \partial_{\mu}\mc{J}^{\mu}_{X}=-\frac{2e}{\pi} E_{x}$ (Note that $Q_{xX}=8\pi/a$).
This is consistent with the chiral modes in different pseudospin sectors that are bound to the screw dislocation.

\bibliography{refs.bib}